\documentclass[structabstract]{aa}  
\usepackage{amsmath}

\usepackage{graphicx}
\usepackage{txfonts}
\begin{document}
\title{The structure of spiral galaxies: radial profiles in stellar 
mass--to--light ratio and the dark matter distribution}

\author{Laura Portinari$^1$ and Paolo Salucci$^2$}

   \institute{$^1$ Tuorla Observatory, Department of Physics and Astronomy,
              University of Turku, V\"ais\"al\"antie 20, FIN-21500 Piikki\"o, 
              Finland\\
              \email{lporti@utu.fi} \\
             $^2$ SISSA/ISAS, via Beirut 4, I-34014 Trieste, Italy\\
             \email{salucci@sissa.it}
             }

   \date{Received November 2008. Accepted April 2009.}

\abstract
{The colour and metallicity gradients observed in spiral galaxies suggest
that the mass-to-light ratio ({\mbox{M$_\star$/L}}) of the stellar disc 
is a function of radius. This is indeed predicted by chemo--photometric
models of galactic discs.}
{We investigate the distribution of luminous and dark matter in spiral 
galaxies, taking into account the radial dependence of the stellar 
{\mbox{M$_\star$/L}}, which is usually assumed to be constant in studies 
of the mass structure.}
{From earlier chemo--photometric models and in agreement with the observed 
radial profiles of 
galaxy colours, we derive the typical average {\mbox{M$_\star$/L}} profile 
of the stellar discs of spiral galaxies.  We computed the corresponding 
variable mass--to--light (VML)
stellar  surface density profile and then  the  VML  disc  contribution  
to the circular velocity.  We used the latter, combined with
a well--studied dark matter velocity profile, to mass model co-added 
rotation curves.}
{By investigating rotation curves in the framework of  
VML stellar discs, we confirm the scenario obtained
with the constant {\mbox{M$_\star$/L}} assumption: a dark matter halo with 
a shallow core, an inner baryon-dominated region, and a larger proportion 
of dark matter in smaller objects.
However, the resulting size of the the dark halo core and of the inner baryon 
dominance region are somewhat smaller. The stronger role that VML discs have 
in the innermost regions is important for constraining the galaxy mass 
structure in both $\Lambda$ cold dark matter and MOND scenarios.}
{}


\titlerunning{Mass distribution of spiral galaxies: considering 
the radial profiles in M$_\star$/L}
\authorrunning{Portinari \& Salucci}
\maketitle
\section{Introduction}
\label{sect:introduction}
Within Newtonian gravity  the mass model of spiral galaxies, including an  
(assumed spherical) dark halo, can be derived from high--quality rotation 
curves with little presence of non circular motions due to bars, spiral 
structure ot oval distortions; or from co-added rotation curves 
(e.g.\ Broeils 1992; Salucci \& Burkert 2000).\footnote{For the scope
of this paper we  use the term  ``rotation curve'' also  in the sense of 
``circular velocity as a function of radius''.} A careful account of the
distribution of the baryonic matter is crucial. This includes
the contribution of the stellar and HI discs, and of a significant 
central bulge in early type spiral galaxies (not investigated here).

The surface density of the stellar disc $\Sigma_\star(R)$ is
obtained from the {\it surface brightness} $I(R)$, often    fitted
with an exponential profile of scalelength $h$ (Freeman 1970)
\begin{equation*}
\label{eq:exp_profile}
\tag{1a}
I(R)= I_0 \, e^{-R/h},
\end{equation*}
by assuming that the stellar mass-to-light ratio 
({\mbox{M$_\star$/L}}) is radially constant:
\begin{equation*}
\label{eq:CML_profile}
\tag{1b}
\Sigma_\star(R)=(M_\star/L) \,\,  I(R),
\end{equation*}
where $I(R)$ is either directly the observed surface brightness or its 
exponential fit. 
Equation~\ref{eq:CML_profile} corresponds to the simplest assumption according
to which
the stellar mass strictly follows  the light. This  is 
reasonable,  especially when  the surface brightness is measured in the red or 
infrared bands that best trace the  stellar mass distribution, in view of the 
 {\it  approximate} radial constancy of  galaxy colours (e.g.\ Kent 
1986; Verheijen 1997). On the other hand, various observational facts suggest 
that  different regions in a spiral galaxy experience different 
evolutionary histories that lead  to different colours and mass-to-light 
ratios. Then,
it is  worth dropping the assumption of a  constant {\mbox{M$_\star$/L}} and 
investigating the effects of realistic radial variations in the 
{\mbox{M$_\star$/L}} ratio on  the mass models.

The main argument for this comes from the observed metallicity gradients: 
in disc galaxies (including the Milky Way) the
metallicity,  traced by the HII regions and by the youngest stars, decreases
outward with a typical gradient of {\mbox{--0.2~dex~$h_B^{-1}$}}
in [O/H] (Garnett et~al.\ 1997; van Zee et~al.\ 1998; Shaver et~al.\ 1983; 
Smartt \& Rolleston 1997; Gummersbach et~al.\ 1998). 
The standard interpretation is that star formation and chemical enrichment 
proceed in the outer regions at a slower pace than in the inner regions,
in an  ``inside--out''  formation scenario  
 (Matteucci \& Fran\c{c}ois 1989; Sommer--Larsen
1991; Ryder \& Dopita 1994; Portinari \& Chiosi 1999 and references therein).  
This  is  supported  by the evidence that
spiral discs are systematically bluer  towards their outskirts and
the scalelength of the light distribution
is typically shorter in red bands than in the blue (de Jong 1996c; Bell
\& de Jong 2000; Jansen et~al.\ 2000). 
Therefore, at a face value, stellar populations of younger age and of lower 
metallicity lie in the outer regions. All this points to a  decline in 
{\mbox{M$_\star$/L}} with radius (Bell \& de Jong 2001), whose extent and 
consequences are ripe for investigation.

The other main player in the mass structure of galaxies is the dark halo. 
This can be modelled by means of a  theoretical approach that stems from the 
favoured $\Lambda$ cold dark matter ($\Lambda$CDM) cosmological 
scenario, very successful at accounting for the large--scale structure of 
the Universe. For virialized objects, this predicts well--defined  
 ``cusped'' profiles (Navarro, Frenk \& White 1996) that, however, have been 
found to be at variance with observations  (e.g.\  Moore 1994; 
Salucci, Walter \& Borriello 2003; de Blok \& Bosma 2002; 
Kuzio de Naray et~al.\ 2006; de Blok 2007; Spano et~al.\ 2008). 
Alternatively, we resort to  an empirical approach: one can easily
find  a form of (cored) DM density distribution that can account for all 
available kinematic data (Gentile et~al.\ 2004, 2005; Salucci et~al.\ 2007).
 
The present work aims to bridge the gap between the chemo--photometric 
studies, which support inside--out formation and  radial gradients
in colours and {\mbox{M$_\star$/L}}, and the mass modelling of spiral galaxies
where {\mbox{M$_\star$/L}} is traditionally assumed to be constant over the 
whole disc (with a few recent exceptions, see below).

Let us briefly review the state of the art of the study of the kinematics 
of spiral galaxies with the aid of chemo-photometric information.
The average {\mbox{M$_\star$/L}} ratio of a  stellar disc, obtained from 
its global broadband colours by means of stellar population synthesis 
techniques, has proved to be a useful tool in dynamical studies 
of  spiral galaxies (Ashman, Salucci \& Persic 1993; 
Bell \& de Jong 2001).
More recently, Kranz, Slyz \& Rix (2003) and Kassin, de Jong \& Weiner 
(2006) 
for the first time have mass--modelled the rotation curve (RC) of a number of 
spiral galaxies, relaxing  the assumption that {\mbox{M$_\star$/L}}(R)= constant
and considering the existence of  (moderate) radial colour gradients.
{From} these they derived the radial variation in {\mbox{M$_\star$/L}} by means 
of the  (allegedly  universal) relationship between the {\mbox{M$_\star$/L}} 
and colour of a mixed stellar population,
\[ < \log \frac{M_\star}{L_K} >= a + b <(B-R)>, \]
found by spectro-photometric models (Bell \& de Jong 2001; Bell
et al.\ 2003).  This new treatment of the stellar disc surface density, 
considered just as a refinement, has seemingly brought out no major 
new features in the mass models of spiral galaxies. 
Presently, colour--{\mbox{M$_\star$/L}} relations are being used in a variety
of contexts where the mass profile of the stellar disc needs to be 
estimated (see e.g.\ Bakos, Trujillo \& Pohlen 2008; Treuthardt, Salo \& Buta
2008).

However, this pioneering way of accounting for variable {\mbox{M$_\star$/L}} 
ratios is still
subject to improvement. In fact, the above relation was derived from the 
{\it global} properties of model galaxies;\footnote{The Bell \& de Jong (2001)
models are detailed multi--zone models accounting for the observed gradients
in colour, metallicity, and age of galactic discs; however, the colour-M/L 
relations they present refer to their global galaxy models. Though similar 
relations have been shown to also hold radially, within individual discs
(PST04) it is not a priori guaranteed that the same relations hold 
globally and locally, 
especially for NIR bands where the age--metallicity degeneracy is lifted. 
This caveat applies in particular to the (much shallower) semi--empirical
relations of Bell et~al.\ (2003), which are the result of 
simple one--zone models calibrated on the {\it global} photometric properties 
of a large sample of SDSS/2MASS galaxies.}
namely, the relation is supposed to hold {\it among} galaxies, relating 
their global colour and mass-to-light ratio. Yet, in the above--cited dynamical
studies, the relation is instead applied {\it within} each object in order  
to estimate the   
{\it  local} {\mbox{M$_\star$/L}} from the local colour $(B-R)$(R). 
Moreover, it is sometimes applied beyond the range of colours for which it was 
established.
Some caution is therefore necessary when such colour--{\mbox{M$_\star$/L}} 
relations are applied radially within individual galaxies: in the inside--out 
scenario, it is likely that radial variations in {\mbox{M$_\star$/L}} are far 
more ``coherent'' than the variations related to global colour from galaxy 
to galaxy.
  
In this paper we adopt a different approach to investigating the  radial
{\mbox{M$_\star$/L}} gradients and their consequences for  the mass structure 
of spiral galaxies.
We work out, for the first time in a self-consistent way, the theoretical 
radial gradients in {\mbox{M$_\star$/L}} of stellar discs by means of 
multi-zone   chemo-photometric models.  These follow the radial history 
of star formation 
  and account for the observed metallicity and colour gradients.
We obtain a typical radial profile of variable mass--to--light ratio (VML)
that, convolved with an exponential light distribution, yields the underlying 
stellar density profile.

We derive the corresponding VML--disc contribution to the circular velocity and
we mass--model the  co-added rotation curves of Persic, Salucci \& Stel (1996),
that represent the average RC of a spiral of a given luminosity 
well (see Salucci et~al.\ 2007 and references therein). 
Then, moving beyond the usual assumption of a  constant 
mass-to-light ratio, we investigate
crucial issues of the mass distribution of disc galaxies. These include  
the ubiquitous presence of DM in every object, the presence of a  region
of ``inner baryon dominance'', the luminosity dependence of the 
dark-to-luminous mass fraction, and finally the inner density profile 
of the dark matter  halo. We also discuss VML discs in the MOND scenario.
    
The paper is organised as follows. In Section~2 we derive the average typical 
{\mbox{M$_\star$/L}} profiles from chemo--photometric models of disc galaxies.
 In Section~3 we show how they 
are supported by the observed colour gradients.
In Section~4 we determine the corresponding VML--disc rotation curve; 
we then discuss the
consequent disc/dark halo decomposition of   rotation curves.
Finally, in Section~5 we draw our conclusions.

\section{Chemo--photometric models and radial gradients
 in stellar mass--to--light ratio}
The standard ``inside--out'' scenario  of disc galaxy evolution was originally 
developed to match a number of chemical properties in the Solar Neighbourhood 
and in the Milky Way (e.g.\ Matteucci \& Fran\c cois 1989; Chiappini et~al.\ 
1997, 2001; Portinari, Chiosi \& Bressan 1998; Portinari \& Chiosi 1999, 2000;
Boissier \& Prantzos 1999). Subsequently, it was successfully used 
to describe the chemical and photometric evolution of individual spiral 
galaxies (e.g.\ Moll\'a, Ferrini \& Diaz 1996, 1997; Renda et~al.\ 2005), 
as well as the whole spiral population (e.g.\ Boissier \& Prantzos 2000, 2001; 
Prantzos \& Boissier 2000; Bell \& Bower 2000; Moll\'a \& Diaz 2005).
In this section we discuss how this scenario necessarily implies that stellar 
discs in spiral galaxies have radial gradients in colour and 
{\mbox{M$_\star$/L}}.

We specifically consider the multi--zone models of Portinari et~al.\ 
(2004; hereinafter PST04) that compute the detailed chemical and photometric 
profiles of the disc and the corresponding radial {\mbox{M$_\star$/L}} 
and colour profiles. Radial profiles were not explicitly presented in PST04 
(which was focused 
on the {\it global} {\mbox{M$_\star$/L} of disc galaxies), so we will do it 
in the present paper. Let us  anticipate (\S\ref{sect:MLprofiles})  
the main result: when tuned to observations,
chemo--photometric models imply that 
the disc  {\mbox{M$_\star$/L} declines with radius $R$ as
\[
\frac{M_\star}{L}(R) \propto \exp \left\{ - b 
\left[ \left( \frac{R}{h} \right)^s -1 \right] \right\} 
\]
with  $0.5 \leq s \leq 1.1$ and $0.14 \leq b \leq 0.8$, and $h$ the scalelength
of the luminous profile. 
Any of the above cited works in the framework of the inside--out paradigm 
lead to similar {\mbox{M$_\star$/L} profiles, though this fact is rarely 
mentioned explicitly (one exception is Fig.~7a of Boissier \& Prantzos 1999),  
probably because the issue lies outside the direct scopes of 
chemical evolution studies.

We briefly  outline the main features of the  code and the relative models, 
but the reader uninterested in the  details can directly go to 
\S\ref{sect:MLprofiles} where we present   {\mbox{M$_\star$/L} profiles 
derived from the PST04 models.
In short: (1) the main observable underlying the derived mass-to-light profile 
is the metallicity gradient of spiral  galaxies 
({\mbox{--0.2 $\pm$ 0.1~dex~$h_B^{-1}$}} in [O/H], quite independent of mass 
and Hubble type); (2) the  PST04 models correspondingly predict   
colour gradients that will be checked in Sect.~3.
 
\subsection{Main features of the models}
Here below we briefly summarise the main features of the chemo--photometric 
models of PST04, but further details can be found in the original paper.
\begin{itemize}
\item
The disc galaxy models possess a multi--zone radial structure, 
divided into (typically 26) concentric annuli, which allows us  to follow 
radial gradients in star formation (SF) history. The local SF and chemical 
evolution history of each annulus are used to compute the radial profiles 
of metallicity, luminosity, and colours.
\item
A galactic disc is assumed to form gradually via slow accretion (``infall'') of
primordial gas, essentially cooling from the gaseous halo and settling into
the disc. An exponentially decreasing infall rate is  invoked to fit 
the metallicity distribution of long--lived stars in the Solar Neighbourhood
(i.e. the ``G--dwarf'' problem; Lynden--Bell 1975; Tinsley 1980; Pagel 1997).
This kind of infall  is also observed in dynamical models of the formation of 
galactic discs
(Larson 1976; Sommer--Larsen 1991; Burkert, Truran \& Hensler 1992; 
Sommer--Larsen, G\"otz \& Portinari 2003).
\item
The present--day surface density distribution of the baryonic matter 
(stars+gas) is forced to an exponential radial profile. With this choice,
the present--day luminosity profile also approaches an exponential, 
as observations indicate, and a scalelength $h$ can be confidently determined, 
at least out to 3--4 scalelengths (see also Boissier \& Prantzos 1999).
\item
The  code  adopts a physically motivated star formation efficiency,  
which increases with the local surface mass density (Dopita \& Ryder 1994;
see Section 6 of PST04). This form of SF law is known to reproduce the 
radial chemical and photometric profiles of galaxies (Portinari \& Chiosi 
1999; Bell \& Bower 2000). 
\item
Once the stellar initial mass function (IMF) is fixed, the main parameters
of the  code  are the SF efficiency $\nu$ (in Gyr$^{-1}$) and the infall 
timescale $\tau_{inf}$. The models are then a two--parameter family. They 
were calibrated to reproduce the {\it typical average} chemical properties
(metallicity and metallicity gradients) of Milky Way--sized 
spiral galaxies, as follows.
A grid of models with different infall timescales $\tau_{inf}$ is developed, 
and for each $\tau_{inf}$, the SF efficiency $\nu$ is tuned 
to reproduce the typical metallicity of spiral galaxies with $M_B \sim -21$.
Specifically, they reproduce 12+log(O/H)=9.1~dex at a galactocentric radius 
$R=1~h_B$ (see Sect.~6.2 of PST04).
Once the absolute values of $\nu$ and $\tau_{inf}$ are so calibrated, 
we consider the {\it radial} metallicity gradient that must match 
the observed {\mbox{--0.2 $\pm$ 0.1~dex~$h_B^{-1}$}} in [O/H] 
(Garnett et~al.\ 1997; van Zee et~al.\ 1998; Prantzos \& Boissier 2000).
This is often obtained with no further tuning of the parameters.   
The adopted SF law in fact, with a dependence on surface density,
can reproduce the observed metallicity profile. Only models with long infall 
timescales reach an agreement  with observations by introducing an additional
radial increase in $\tau_{inf}(R)$  
($\tau_{inf}(R) \propto R^x,~0.5 \leq x \leq 2$, see Table~2 of PST04).
\item
We used  six different IMFs: those of Salpeter (1955), Kroupa (1998), 
Kennicutt (1983), Larson
(1998), and Chabrier (2001, 2002).  The other IMFs are called
``bottom--light'' with respect to Salpeter, in that  they store less mass 
in the low--mass end of the distribution ($<1 \, M_{\odot}$) than the 
Salpeter IMF (extended down to 0.1~$M_{\odot}$). These IMFs are 
favoured by direct estimates of the stellar mass function 
in the Solar Neighbourhood, in the Galactic bulge, and in globular clusters
and are indirectly  supported  by Tully--Fisher arguments
(PST04). The upper mass limits on the IMFs were,
in some models, tuned to reproduce the observed gas fractions of late--type 
disc galaxies, but such fine tuning has very  little relevance for the aims 
of this paper.
\end{itemize}
We thus obtained, for each of the IMFs considered, a series of models with 
($\nu, \tau_{inf}(R)$) calibrated to reproduce the observed metallicity of 
Milky Way--sized spiral galaxies, and the typical
metallicity gradients (which are quite independent of mass and Hubble type).

Increasing infall timescales $\tau_{inf}$'s ``produce'' objects with more 
extended SFHs (larger ``birthrate parameters'', the ratio between present--day 
and past average SFR) and  bluer colours, that represent galaxies 
of later  Hubble types (Kennicutt, Tamblyn \& Congdon 1994).

Finally, we  stress that,  while the absolute value of metallicity 
is a function of galaxy mass, the metallicity gradient is quite independent 
of mass and Hubble type (Garnett et~al.\ 1997; van Zee et~al.\ 1998; 
Prantzos \& Boissier 2000). 
Therefore, the  {\it radial trends}  obtained for the Milky Way--sized galaxies
modelled by PST04 are actually valid in general.

Barred galaxies may be an exception, as they often present much shallower 
metallicity gradients (e.g.\ Martin \& Roy 1994), 
probably a result of bar--induced radial gas 
flows that wash out any pre-existing gradient.
However, this phenomenon is probably limited to strong bars;
for instance, our Milky Way hosts a bar (or two; Lopez--Corredoira et~al.\ 
2007, and references therein) yet its metallicity
gradient is in line with the one we model --- and has actually 
inspired the whole inside--out scenario since the 80's. Besides, bar--induced
gas flows wash out only the present--day metallicity gradient in the gas phase,
while the underlying metallicity, colour, and {\mbox{M$_\star$/L}} 
gradients in the stellar component are much less affected,
so our models should reasonably describe barred galaxies, as long as 
their colour gradients are similar to those of other galaxies.

Clearly, our present discussion does not apply in individual objects 
where no significant metallicity and colour gradients are observed
(whether due to bar mixing or other reasons); however, this paper focuses 
on average trends and we show in 
Section~\ref{sect:colours} that the models yield reasonable predictions 
for the average colour gradients of disc galaxies.
For the sake of completeness, we mention that radial mixing of stellar 
populations may also affect the chemo-photometric evolution of disc galaxies
(Ro\v{s}kar et~al.\ 2008; Sch\"onrich \& Binney 2009). While very promising 
for understanding the detailed connections between kinematics and chemistry
in Galactic stars, models including these effects are still in their infancy.
Whether they represent just a refinement over the underlying inside--out 
scenario or will drastically change the standard interpretation of the 
metallicity, colour, and {\mbox{M$_\star$/L}} gradients of disc galaxies
remains to be explored, especially considering that such models involve many 
more free parameters than the standard ones. In this paper we necessarily limit
our discussion to the dynamical consequences of the decade--long, 
well--established inside--out scenario.

\subsection{The radial profile of {\mbox{M$_\star$/L}}}
\label{sect:MLprofiles} 
It is well known that the  absolute value of {\mbox{M$_\star$/L}} 
depends on the SF history, stellar metallicity distribution function,
and underlying IMF. Global information, such as absolute values of the
metallicity and colour, is very valuable to frame the galaxy properties
and compare galaxies to each other ---
though is {\it per se} unable to constrain the mass models with a precision
comparable to kinematic analysis (see Salucci, Yegorova \& Drory 2008).

However, in this paper we are not interested in differences among different 
objects, but focus on the internal radial profiles in 
{\mbox{M$_\star$/L}} and their dynamical consequences compared to the simple
constant {\mbox{M$_\star$/L}} assumption.
Therefore, we shall adopt a suitable scaling that cancels out object-to-object
offsets and highlights the radial {\mbox{M$_\star$/L}} variation within the 
discs.

We define the $I$ band scalelength $h_I$ by fitting
the model $I$ band luminosity profile with the exponential law
of Eq.~\ref{eq:exp_profile} in the range $0.5-3~h_I$. Introducing the notation 
{\mbox{M$_\star$/L}}$_{\lambda} (R) \equiv \Upsilon_{\lambda,*} (R)$, 
we take the value of the {\mbox{M$_\star$/L}} at one scalelength, 
$\Upsilon_{\lambda,1}=${\mbox{M$_\star$/L}}$_{\lambda}(h_I)$
as the zero--point normalization of the {\mbox{M$_\star$/L}} and
we define  the normalized {\mbox{M$_\star$/L}} radial profile as
\begin{equation*}
\label{eq:ML_scaled}
\tag{2}
\frac{\Upsilon_{\lambda,\star} (R)}{\Upsilon_{\lambda,1}} =
\frac{\frac{M_\star}{L_{\lambda}}(R)}{\frac{M_\star}{L_{\lambda}}(h_I)}.
\end{equation*}
The normalization eliminates the zero--point 
offsets related to the specific IMF and star formation history of each model. 
The normalized {\mbox{M$_\star$/L}} profiles of the models show very similar
behaviour, with representative examples shown in Fig.~\ref{fig:MLprofiles}.
The profiles range from ``shallow'' (close to linear out to $\sim 3 h$ 
and shallower farther out; top panel) to ``concave'' (bottom panel).  
 
In any case, {\mbox{M$_\star$/L}} decreases with radius in all models, simply 
as an effect of
the age and metallicity gradients of the stellar populations that  evolve in an
inside--out scenario.  Quantitatively, {\mbox{M$_\star$/L$_I$}}  decreases by 
about a factor of 2
between 1 and 3--4 scalelengths, with a rise towards the centre for the concave
cases. Though the radial gradients in {\mbox{M$_\star$/L}} are stronger in 
bluer bands, they
are also non--negligible in the $K$ band: {\mbox{M$_\star$/L$_K$}} decreases 
by about 40\% between
1 and $4 \, h_I$. As a result, the stellar surface  density
profile is generally  more  concentrated than the brightness profile, 
even in the NIR (see also Kranz et~al.\ 2003).
 
To proceed, we first consider that the $I$ band offers a number of 
observational and theoretical advantages (see PST04), so that we choose it 
as the reference band --- as in the kinematic analysis
of \S~\ref{sect:decomposition}. 
We find that the normalized {\mbox{M$_\star$/L$_I$}}  profiles are all  
approximated well by
\begin{equation*}
\label{eq:ML_profile}
\tag{3a}
\frac{\Upsilon_{I,*}(R)}{\Upsilon_{I,1}} = \exp \left\{ - a(s) 
\left[ \left( \frac{R}{h_I} \right)^s -1 \right] \right\} 
\end{equation*}
with
\begin{equation*}
\tag{3b}
  a(s) =\frac{5}{4} \, (1.3-s)^3 + 0.13 .
\end{equation*}
In detail, shallow {\mbox{M$_\star$/L$_I$}} profiles have $s \simeq 1$ 
(0.9--1.1), i.e.\ close to exponential, while concave profiles 
are well--fitted by  $s=0.5-0.7$; e.g., the example concave profile 
in the bottom panel of Fig.~\ref{fig:MLprofiles} corresponds to $s=0.5$. 
The standard constant mass--to--light assumption is recovered for $s=0$.

The analytical fit in Eq.~3 is accurate to 2--3\% at $R \geq 0.5~h$ and 
to about 10\% in the very central regions. 
For the typical "average"  spiral  galaxy the actual value of $s$ 
will lie within the (reasonably narrow) range of values discussed above. 
In passing, let us point out that it is up to future investigations to test 
whether Eq.~3 can be also representative of {\it individual} objects, 
and determine the corresponding values of $s$.

\begin{figure}
\centering
\includegraphics[width=9truecm]{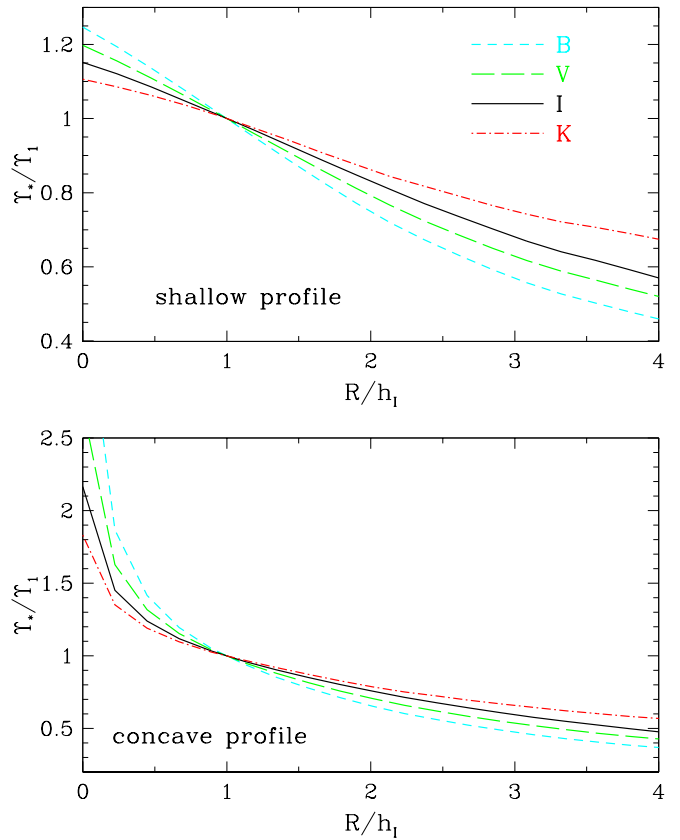}
\caption{ {\mbox{M$_\star$/L}} profiles in different bands predicted 
by our chemo--photometric 
models, scaled to the respective values at $R=h_I$ 
(Eq.~\protect{\ref{eq:ML_scaled}}). 
An example of {\mbox{M$_\star$/L}} with shallow profile (model {\sf Kenn-H} 
of PST04, 
top panel) and one with concave profile (model {\sf Chab-F}, bottom panel) 
are shown.}
\label{fig:MLprofiles}
\end{figure}

\begin{figure}
\centering
\includegraphics[width=9truecm]{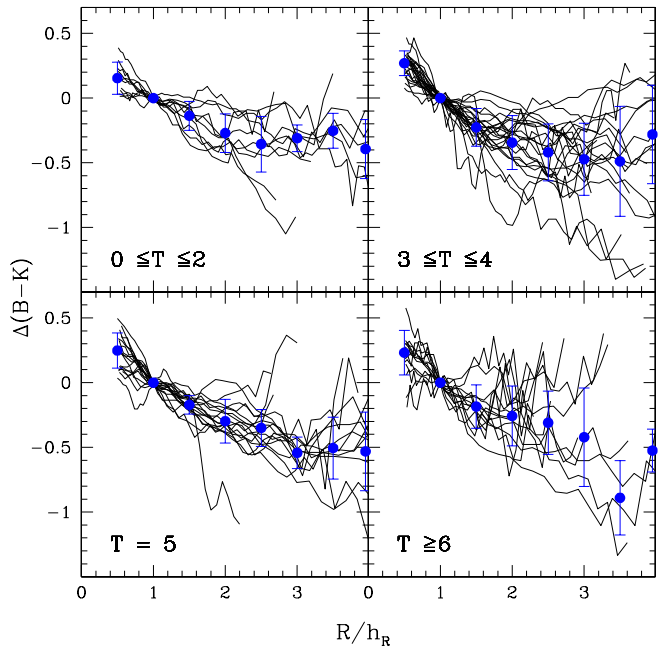}
\caption{Colour difference profiles for the sample of disc galaxies of de Jong
\& van der Kruit (1994).  T codes the galaxy morphological type: T=0--2
(Sa--Sab), T=3--4 (Sb--Sbc), T=5 (Sc), T=6--10 (Scd and later types).
Dots with error bars represent the average $\Delta({\rm colour})$ and 
its 1~$\sigma$ dispersion for the co-added profiles.}
\label{fig:colprofiles_T}
\end{figure}

\section{Colour gradients}
\label{sect:colours}
It is well known that spectro-photometric models of spiral galaxies predict 
 tight relations between the {\mbox{M$_\star$/L}} and the colours 
of (simple and composite) stellar populations (Bell \& de Jong 2001; PST04).
Henceforth, the observed colour gradients in spiral galaxies can be considered 
the empirical counterpart of the {\mbox{M$_\star$/L}} gradients.
An obvious test for our model {\mbox{M$_\star$/L}} profiles is then 
to compare the corresponding colour profiles, predicted to get bluer outwards,
to observations. 
To this purpose we consider the 
multi--band sample of 86 disc galaxies of de Jong \& van der Kruit (1994).
It is worth noting that the {\it overall} colour of a spiral galaxy tends to be
redder for earlier morphological types (e.g.\ Roberts \& Haynes 1994) because
of their different average  disc SF histories, rather than to systematic 
variations in the bulge--to--disc ratios (Kennicutt et~al.\ 1994; de Jong
1996bc). To circumvent this intrinsic variance in the global colour of
different galaxies, and highlight the {\it radial} trends within each of them, 
we normalize the colour data by taking the colour measured at 1 scalelength $h$
as the zero--point and by shifting all the values as
\begin{equation*}
\label{eq:colour_shift}
\tag{4}
\Delta({\rm colour}) = {\rm colour}(R) - {\rm colour}(h) .
\end{equation*}
This  shift is consistent with the {\mbox{M$_\star$/L}} scaling 
of Eq.~\ref{eq:ML_scaled} that  we apply to the models.

We show the resulting ``colour difference profiles'' in
Fig.~\ref{fig:colprofiles_T}, which is obtained from Fig.~2 of de Jong 
(1996c) after the zero--point shift by expressing 
the radial coordinate in disc scalelengths (taken from de Jong 1996a). 
For the sake of example we show the run of  $B-K$ colour versus radius 
measured in $R$--band scalelengths, but similar trends hold in  
other bands (see also Fig.~\ref{fig:colprofiles_models}). 
We ignore the data of the innermost half scalelength, which are heavily 
affected by the bulge component.  
Fig.~\ref{fig:colprofiles_T} clearly highlights 
the existence of negative radial colour gradients, out 
to at least 3 scalelengths. For different morphological types
the colour  gradients are quite similar in the range $0.5 \leq R/h \leq 3$, 
with Scd and later type galaxies showing a larger scatter.
Beyond $3 h$ the colour profiles become shallower (and noisy).

In spite of the considerable scatter (and of uncertainties in sky background 
subtraction at the outermost radii), there is no question that the colours 
{\it on average} become bluer outwards; $B-K$, for example, decreases by about 
$0.25~mag$ per disc scalelength. {From} this, we can claim that colour 
gradients emerge as a systematic, constitutive property of galactic discs. 

Then, in Fig.~\ref{fig:colprofiles_models} we overlay the colour profiles 
of all morphological types $0 \leq T \leq 6$ (neglecting later, more irregular
types) and we compare their average values with the predictions of our 
chemo--photometric models.
In detail, we plot various colour profiles in different bands and 
compare the data with the model predictions obtained by assuming 
a variety of IMFs. 
Each panel shows, for a given set of IMF models, two examples of model
colour gradients, the shallower/steeper one corresponding
to the shallow and concave {\mbox{M$_\star$/L}} profiles in 
\S\ref{sect:MLprofiles}.

For all the adopted IMFs, a good agreement results between the colour profiles
predicted by the models for an ``average'' spiral and the actual observations.  
Therefore, in the inside--out formation scenario of stellar discs, colour
gradients emerge naturally and are remarkably similar to the observed ones.

Notice that we neglect  the innermost $0.5 h$ region, where the
comparison would be inappropriate, as data is affected by a bulge
component,  not included in  our models. It is not our aim to study 
bulge--dominated discs
 or the bulge--disc interface, and since the innermost $R < 0.5 \, h$ 
region encloses only 10--15\%
of the disc mass, the lack of a more exact treatment does not significantly 
bias our investigation on VML--discs.

In conclusion, the observational properties of stellar discs 
indicate that their {\mbox{M$_\star$/L}} is likely a function of radius
and that chemo--photometric models predict the average radial colour 
and {\mbox{M$_\star$/L}} profiles of spiral galaxies reasonably well. 
 
\begin{figure*}
\centering
\includegraphics[angle=-90,width=19.5truecm]{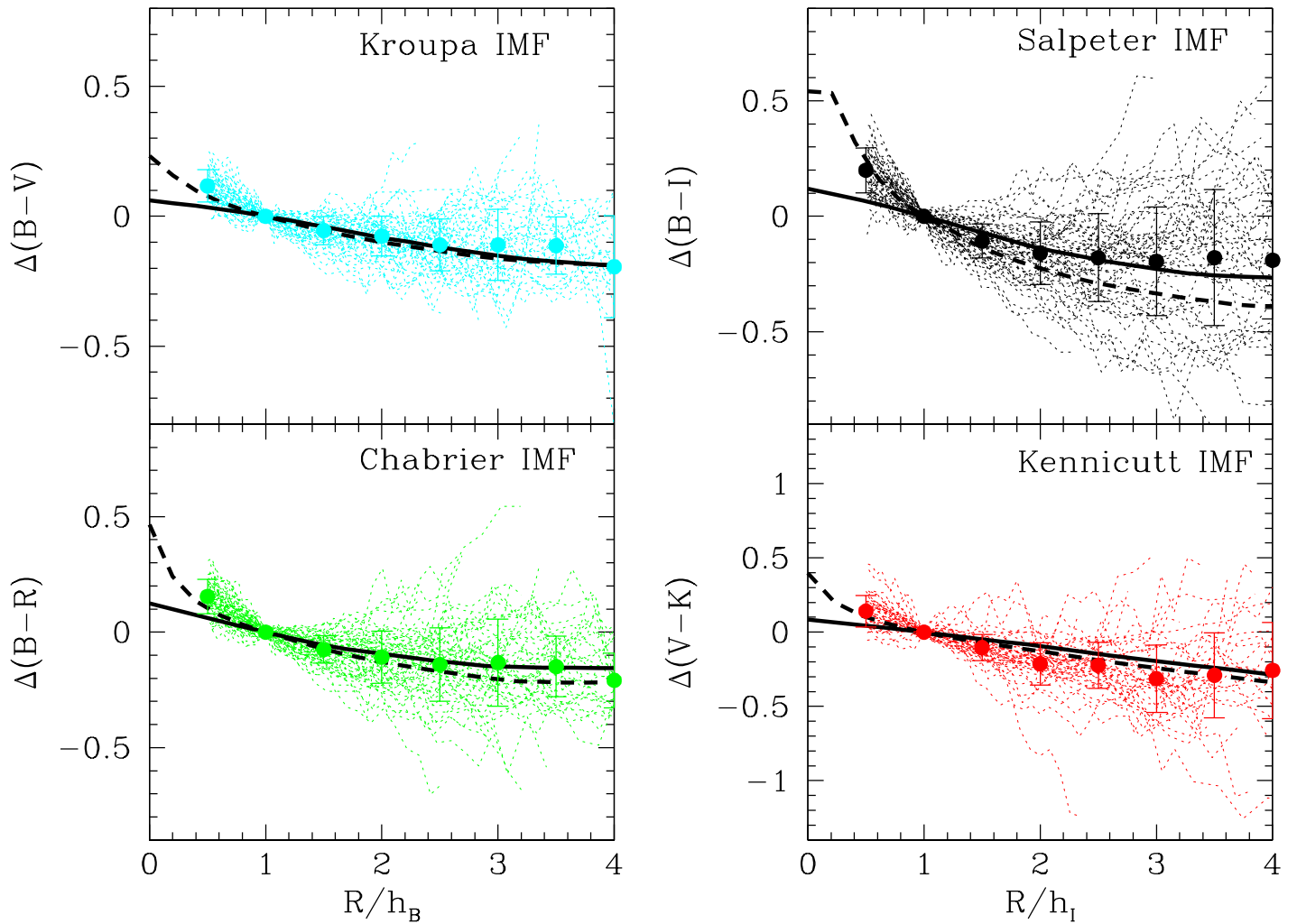}
\caption{Radial colour difference profiles in different bands, computed
relative to the colour at $R=h$ (Eq.~\protect{\ref{eq:colour_shift}}).
{\it Thin dotted lines}: data from de Jong \& van der Kruit (1994) 
for $0 \leq T \leq 6$; 
{\it dots with error bars}: corresponding average $\Delta({\rm colour})$ 
and 1~$\sigma$ dispersion;
{\it thick solid and dashed lines}: chemo-photometric models from PST04
(specifically, the models plotted are {\sf Krou-C,E; Salp-A,E; Chab-A,F; 
Kenn-H,J}).}
\label{fig:colprofiles_models}
\end{figure*}

\begin{figure*}
\centering
\includegraphics[angle=-90,width=19.5truecm]{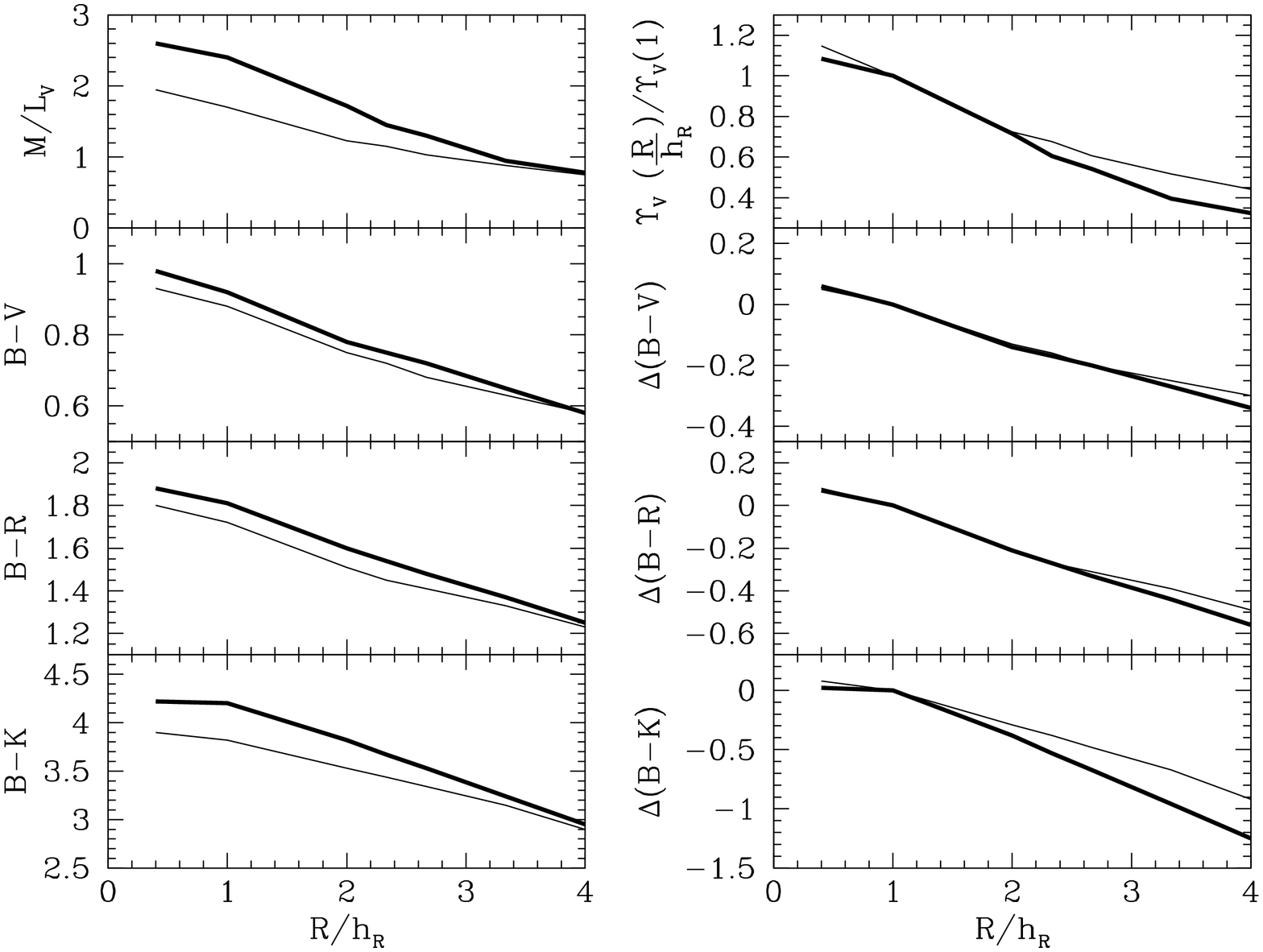}
\caption{Radial {\mbox{M$_\star$/L$_V$}} and colour profiles from the Milky Way
model of Boissier \& Prantzos (1999); {\it thin lines}: dust--free predictions;
{\it thick lines}: including extinction. The right panels show the relative
{\mbox{M$_\star$/L$_V$}} and colour profiles, scaled as discussed in
\S~\protect{\ref{sect:MLprofiles}} and~~\protect{\ref{sect:colours}}.}
\label{fig:dust}
\end{figure*}

\subsection{Dust effects}
In our models and in the comparison with the observed colour profiles,
we have neglected the effects of dust. This is to a first approximation
justified, since the observational sample consists of face--on galaxies 
for which dust has been found to play a minor role in the colour {\it 
gradients} (de Jong 1996c). 
For an extended sample of low inclination galaxies, Bell \& de Jong (2000) 
similarly conclude that colour profiles are driven by gradients 
in the age and metallicity of the stellar populations; dust induces
a much smaller additional gradient. 
Besides, again to first approximation, dust does not alter the 
colour--{\mbox{M$_\star$/L}} relation (the reddening--extinction dust vector 
running almost parallel to the relation, see Bell \& de Jong 2001). Therefore
the {\mbox{M$_\star$/L}} profiles corresponding to the observed colour profiles
are also expected to be mainly driven by stellar population gradients 
rather than by dust effects.

On the theoretical side, the chemo--photometric model of 
Boissier \& Prantzos (1999) for the Milky Way also predicts that dust 
slightly enhances, but does not create colour gradients. As their models 
are quite similar to those of PST04, we take the results from their Fig.~7 
to illustrate the point. In Fig.~\ref{fig:dust} we show the 
{\mbox{M$_\star$/L$_V$}} and the colour profiles of the Milky Way model
of Boissier \& Prantzos (1999): dust--free predictions and (face--on) 
extinction.
The left panels reproduce their Fig.~7, with the radial coordinate
changed from $R$ to $R/h_R$, where $h_R \sim 3$~kpc (estimated 
from the surface brightness and colour profiles in their Fig.~7).
The $R$--band scalelength should be quite close to the $I$--band scalelength, 
which we refer to in our discussion of the {\mbox{M$_\star$/L}} gradients and 
rotation curves.
The right panels show the same {\mbox{M$_\star$/L$_V$}} and colour
profiles, but with the {\it relative} scalings adopted
in Sects.~\ref{sect:MLprofiles} and~\ref{sect:colours}. Clearly 
the absolute values of colours and {\mbox{M$_\star$/L}} are affected by dust
(left panels), but in relative terms, the effect on the colour gradients 
is small with respect to the amplitude of the gradient. The strongest effect 
is on optical--NIR colours like $B-K$, as noticed by de Jong (1996c).
Comparison to Fig.~\ref{fig:colprofiles_models} shows that the effect
of dust on the colour gradients is much less than the scatter among
individual galaxies, and it does not significantly change the discussion
in the previous section.
As to {\mbox{M$_\star$/L$_V$}}, dust enhances the gradient but remains 
a second--order effect with respect to the main trend set by the inside--out
scenario; and extinction affects the $I$--band (relevant for the remainder
of our paper) much less than the $V$--band. 

In summary, dust certainly plays a non--negligible role when discussing 
the absolute values of colours, {\mbox{M$_\star$/L}} ratios, and age 
of the underlying stellar populations, but its role is secondary
in terms of the gradients discussed in this paper.
Dust effects certainly deserve further attention, especially 
for inclined galaxies, but they will not change our main point.
Radial {\mbox{M$_\star$/L}}
gradients due to stellar populations must exist according to the 
inside-out scenario, and they significantly change the disc rotation curve 
with respect to the classic Freeman one (Sect.~\ref{sect:VML_curve}). 
From Fig.~\ref{fig:dust}, we expect that dust at most {\it enhances} 
the trends we discuss.
%

\section{Variable mass--to--light  rotation curve}
\label{sect:VML_curve}
To establish the actual importance of the radial variations of
the stellar {\mbox{M$_\star$/L}},  a complete chemo--photometric  investigation 
of a large sample of individual objects would be needed, which is beyond 
our present scope. Here we aim to investigate the effects of the average 
stellar {\mbox{M$_\star$/L}} gradient of spiral galaxies (Eq.~3), in 
connection with their average rotation curve (RC), defined as the co-added 
curve of a suitably 
large sample of RC's of objects of fixed luminosity.

{From} now on, we specifically consider the $I$ band, which is the reference 
band for the set of rotation curves studied in 
\S~\ref{sect:decomposition} and  drop the subscript $I$ for simplicity.
{From} Eq.~3 for the {\mbox{M$_\star$/L}} profile we can build the average 
luminous component of the gravitational potential of spiral galaxies. 
We consider values of $s$=0.5, 0.7, and 1.0 (Fig.~\ref{fig:MLIprofiles_as}), 
which cover the range of profiles obtained from the models.
 
We assume that the Freeman profile (Eq.~\ref{eq:exp_profile}) 
describes the light distribution. 
The (small) deviations we observe in the surface brightness of some 
spirals are important for individual mass modelling but  irrelevant for the
present purposes.
As a consequence of the radial gradients in {\mbox{M$_\star$/L}}, the stellar 
surface density 
$\Sigma_\star(R)$  is somewhat more concentrated than the surface 
luminosity profile.
For an exponential disc (Eq.~\ref{eq:exp_profile}),  from Eq.~3 we get
\begin{equation*}
\label{eq:Sigma_profile}
\tag{5}
\Sigma_\star(R) =\Sigma_1 \exp \left\{ - a(s) \left[ \left( \frac{R}{h}
\right)^s -1 \right] - \left[ \left( \frac{R}{h}
\right) -1 \right] \right\}
\end{equation*}
where $\Sigma_1=\Sigma_\star(h)=\Upsilon_1 \, I_0/e $ is the surface density at
1~scalelength. For a given $\Sigma_1$, the surface density has a central value
$\Sigma_\star(0)=\Sigma_1 e^{a+1}$, higher than that of the 
{\mbox{M$_\star$/L}}=constant case by a
factor $e^a$ ($\simeq 1.2-2$ for $s$ in the range 0.5--1.1), 
and decreases more rapidly outward than in the standard
mass--follows--light assumption.  In the case of $s=1$, the surface
density still has an exponential profile, but with scalelength $h/(1+a)$,
i.e.\ about 20\% shorter than that of the surface brightness profile. In this
case we note that the main  VML feature, a mass scalelength slightly less 
than the light scalelength, may have been so far masked by the intrinsic
difficulties in deriving the surface brightness scalelength $h$ well within the
20\% precision level.

In Fig.~\ref{fig:MLIprofiles_as} we show three analytical {\mbox{M$_\star$/L}} 
profiles with $s$=0.5, 0.7, and 1.0, and in Fig.~\ref{fig:newMdisc} we show 
the corresponding cumulative disc masses compared to the constant
{\mbox{M$_\star$/L}} (CML) case. The
radial mass build-up is obviously more rapid and saturates more quickly in the
VML case, with up to  70\% of the total mass contained within about 2~$h$ 
and up to  90\% of the mass within $R_{opt}=$3.2~$h$.

By solving the appropriate Poisson equation we compute numerically the
self--equilibrium circular velocity of thin discs with a VML surface density as
given in Eq.~\ref{eq:Sigma_profile}. In Fig.~\ref{fig:newVrot} we show the 
resulting rotation curves
for the three representative VML discs of Figs.~\ref{fig:MLIprofiles_as}
and~\ref{fig:newMdisc}, compared to the standard CML
rotation curve. The VML RC rises more steeply and peaks 
at smaller radii (1.5--1.9~$h$) than the Freeman CML rotation curve (2.15~$h$),
and it falls off also
more rapidly after the peak.  The VML RCs approach the corresponding Keplerian
fall--off (defined as the region where the logarithmic slope of the RC is close
to --0.5) already at $R=3-3.5 \, h$, rather than at $R> 4 h$ as it occurs for 
the CML disc; for $R \simeq 5 \, h$ the VML RCs finally overlap with the CML 
RC.

It is useful to provide, for the general VML RCs, an analytical fit out to 
3.5~$h$:
\begin{equation}
\label{eq:VML_RC}
\tag{6}
V_{VML}^2(R) = V_{VML}^2 (h) \, n(s) \,
\frac{\xi^{0.9+0.4 s}}{(\xi^2+(1.35+0.5 s)^2)^{1+0.3 s}}
\end{equation}
with $\xi= R/h$ and $n(s)=(1+(1.35+0.5 s)^2)^{1+0.3 s}$ 
a normalization factor.
The fit is precise to better than 2\% between $0.3 \leq \xi \leq 3.5$, and
better than 10\% down to $\xi \geq 0.15$. At larger radii, the VML RCs 
are described well by a Keplerian fall--off.  
The relation between the total disc mass
and the value of the circular velocity at $R_m=2.15~h$, the radius where the 
CML disc RC has its maximum, reads as
\[ M_D= \eta \, G^{-1} V^2(R_m) R_m \]
with $\eta = (0.98, 1, 1.03)$ for the cases of $s= (0.5, 0.7, 1)$, and 
$\eta = 1.2$ for the Freeman CML disc.

\begin{figure}
\centering
\includegraphics[angle=-90,width=8.5truecm]{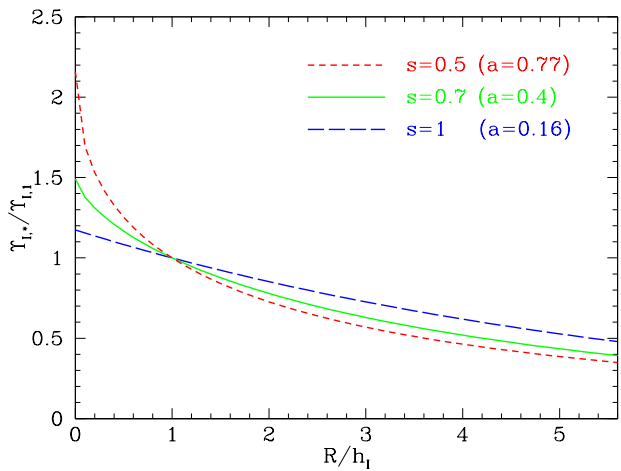}
\caption{Three representative examples of analytical {\mbox{M$_\star$/L$_I$}} 
profiles derived from the chemo--photometric models (Eq.~3).}
\label{fig:MLIprofiles_as}
\end{figure}

\subsection{The dark--luminous matter interplay}
\label{sect:decomposition}
The  stellar surface density $\Sigma_\star(R)$ in spiral galaxies is then only
exponential to a first approximation; more in detail, it is an
exponential modulated by the 
{\mbox{M$_\star$/L}} gradient, as given by Eq.~\ref{eq:Sigma_profile}. 
$V_{VML}(R)$ is somewhat different from the standard CML rotation curve.
The difference is moderate for the ``average'' spiral as defined in this work, 
but in individual objects, it may not be so. {From} the wide range of colour 
profiles observed in individual galaxies (e.g.\ Figs.~\ref{fig:colprofiles_T}
and~\ref{fig:colprofiles_models}), we expect cases in which the VML effect is
more relevant than the average case considered here, while 
it will be negligible in other cases. 

Here, we take the function in  Eq.~\ref{eq:VML_RC} as the typical average 
contribution to the circular velocity of VML stellar discs. We combine it
with a simple halo distribution, to mass--model the typical RCs of disc 
galaxies. Our main aim here is to compare the corresponding disc/halo
decomposition to the results of the standard CML assumption; namely, we perform
a differential analysis, rather than making absolute claims on 
disc fractions, halo profiles, etc.

We consider the co-added rotation curves of Persic, Salucci \& Stel (1996; 
hereinafter PSS96). These are obtained by binning a very large number of 
individual high surface--brightness, late type, bulge-free RCs
(see also Salucci et~al.\ 2007 for references on independent work 
on this issue. More information can be retrieved 
at  http://www.facebook.com/home.php\#!/group.php?gid=3102604\\
50630).
 One advantage of testing a template of co-added RCs 
with a template VML disc--velocity contribution is that this allows 
weak systematic VML kinematic features to be emphasized also, which 
in individual RCs would be below the noise level created by  modelling 
uncertainties and/or observational errors. 
On the other hand, relevant VML features occurring in a fraction of 
objects with strong colour gradients can only be studied in individual RCs.
Ultimately, it is in the latter that the proof for the existence 
of VML discs must be found.  

\begin{figure}
\centering
\includegraphics[angle=-90,width=8.5truecm]{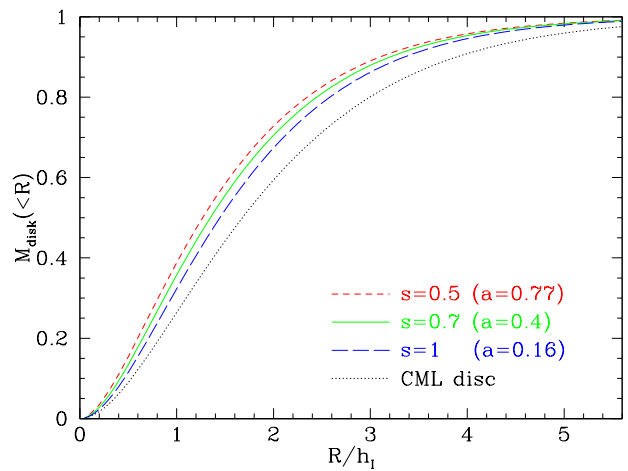}
\caption{Cumulative disc masses corresponding to the {\mbox{M$_\star$/L$_I$}} 
profiles in Fig.~\protect{\ref{fig:MLIprofiles_as}}, compared to the constant
mass--to--light (CML) assumption.}
\label{fig:newMdisc}
\end{figure}

\begin{figure}
\centering
\includegraphics[angle=-90,width=8.5truecm]{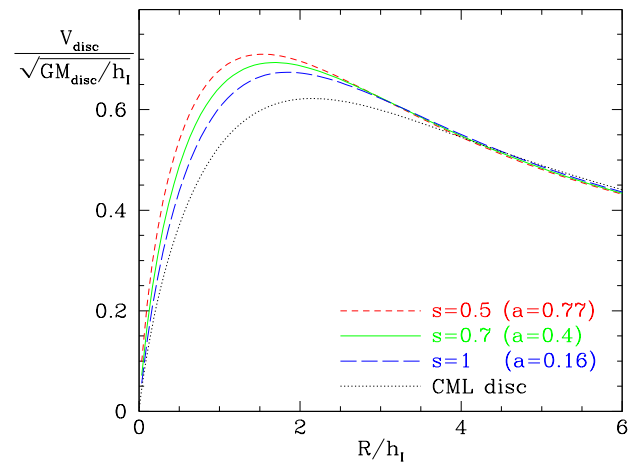}
\caption{Three representative examples of new VML disc rotation curves ({\it
solid and dashed lines}) compared to a constant {\mbox{M$_\star$/L}} 
(CML) Freeman disc
({\it dotted line}). All velocities are normalized to the respective total disc
mass and $I$ band photometric scalelength.}
\label{fig:newVrot}
\end{figure}

\begin{figure*}
\centering
\includegraphics[angle=-90,width=18.5truecm]{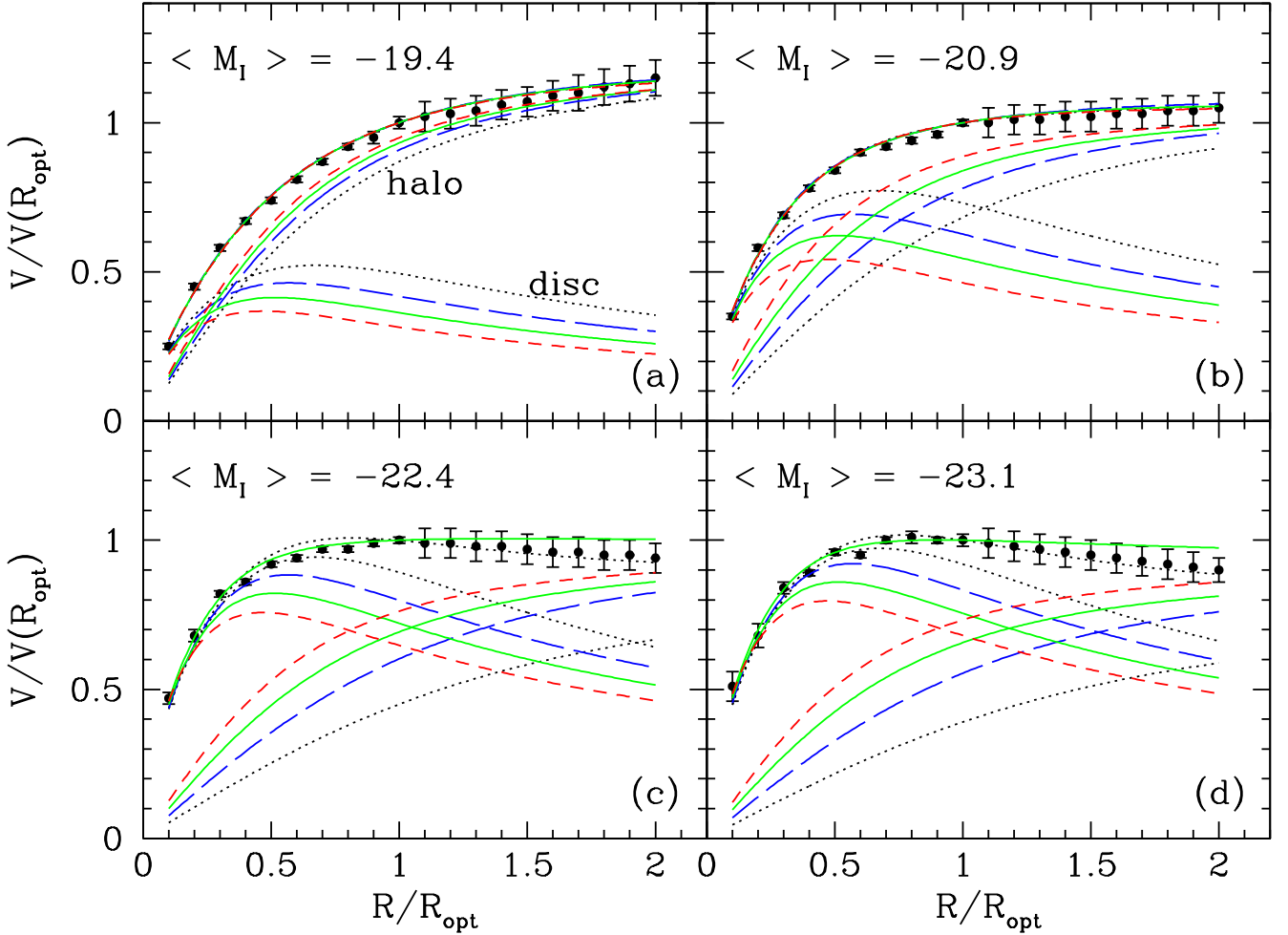}
\caption{Disc/halo decomposition of the co-added rotation curves of PSS96 for
different galaxy luminosities; the disc RC, the dark halo contributions, and 
the total RC are shown for each choice of VML (or CML) disc RC. Line 
symbols are as in Fig.~\protect{\ref{fig:newVrot}}; the dotted lines represents
the PSS96 mass model with the CML disc curve. In panels {\it (c,d)} 
the total (disc+halo) RCs of the three VML disc decompositions overlap.}
\label{fig:decomp}
\end{figure*}

We assume the simplest possible dark--halo velocity contribution (see PSS96):
\begin{equation*}
\tag{7}
{V_{halo}}^2 = V_{opt}^2 (1-\beta) (1+\alpha^2) \frac{x^2}{(x^2+\alpha^2)}
~~~~~~x = \frac{R}{R_{opt}} = \frac{\xi}{3.2} 
\end{equation*}
with $R_{opt} = 3.2 \, h_I$, $V_{opt} $ the {\it observed} circular velocity at
$R_{opt}$ and $(1-\beta)^{1/2}$ the fractional contribution to $V_{opt}$ of the
halo component. Correspondingly, the disc contribution (whether VML or CML)
is normalized so that ${V_{disc}}^2(R_{opt}) = \beta V_{opt}^2$.

Although simple, the function in Eq.~7 has several merits. First, inside $4~h$,
it represents the NFW velocity profile for $\alpha \leq 0.2$. 
Moreover, for appropriate choices of the parameter  $\alpha$, 
it also mimics the pseudo-isothermal and the Burkert (1995) halo velocity 
profiles. 
Then, it is worth recalling that, for any value of its structural parameters 
($\alpha$ and $\beta$), the slope of our halo velocity law (and of the 
profiles quoted above) is very different from that of the  disc contribution. 
At any radius we have
\[ \left| \frac{d\log V_{halo}(R; \alpha, \beta)}{d\log R} - 
\frac{d\log V_{disc}(R; \beta)}{d\log R} \right| >  
\Delta \left| \frac{d\log V(R)}{d\log R} \right| \]
where the righthand side is the uncertainty with which we measure the RC 
slope. This ensures that, in the case a) of  high-quality  rotation curves 
($ \Delta \, | d\log V(R)/d\log R)| <0.04$)  and b) we know the actual 
{\it distribution}  of the luminous matter, 
the mass model we obtain by $\chi^2$ fitting the RC is essentially 
unique. Values of $\alpha$ and disc mass significantly different from the 
best-fit ones, imply mass models that significantly fail to fit the 
observed rotation curve.       
In the mass modelling we neglect the gas contribution to the RC since
it is always negligible out to $\sim 3 \, h$ --- the region 
where the differences between VML and CML RCs are maximal.

We fit the co-added RCs $V_{coadd}(x)$ with our 2-free parameters velocity 
model: $V_{model}^2= V_{halo}^2(x; \alpha, \beta) + V^2_{VML}(x;\beta)$, 
with $V_{VML}$ given by Eq.~\ref{eq:VML_RC}. 
Let us recall that the co-added RCs we adopt here have been modelled in PSS96  
with the same halo velocity profile of Eq.~7 plus a CML 
exponential thin disc. The mass models obtained there (shown in 
Figure~6 and in Eq.~11a,b of PSS96) will serve as a gauge for the models 
we obtain with the present investigation.  
 
We perform $\chi^2$ analysis: the fitting parameters ($\alpha$ and $\beta$)
are optimized  with a Leverberg-Marquardt algorithm (Press et~al.\ 1992).
The increase in the r.m.s.\  scatter in the velocity data 
beyond $R_{opt}$ comes from the fact that, while out to $R_{opt}$ the template 
velocities come from the co-addition of 616 individual curves, at outer radii 
they are 
obtained in a different, non-trivial way (see Fig.~3 and Section 4 of PSS96). 
Then, the related  r.m.s.\ scatter is larger and consequently the velocity 
data for $R>R_{opt} $ do not contribute much toward setting the best--fit 
velocity model.    

The resulting best--fit solutions for 3 different VML cases, which progressively
depart from the CML one, are given in Table~1 and in Fig.~\ref{fig:decomp},
compared to the CML solution.
As a result of the true universality of the RCs of spiral galaxies,  
the r.m.s.\  errors of the co-added curves are very small: 
$\Delta \ V/V  < 0.04 $ (for $R<3 \, h$). 
Within the 2--component halo + disc model, this guarantees that the 
$1 \sigma$ fitting uncertainties on the  parameters of any velocity  
model that fits the data well will be also small. In fact, we find that 
the $1 \sigma$ contour fitting uncertainty on $\alpha$ and $\beta$ is 
between 15\% and 30\%, not surprisingly similar to the corresponding 
uncertainties of the CML + halo mass model in PSS96.

\begin{table}
\caption{Parameters $\alpha$ and $\beta$ and inner baryon dominance radius
of the disc/halo decompositions in Fig.~\protect{\ref{fig:decomp}}, 
for CML and VML disc RCs.}
\label{tab:decomp}
\centering
\begin{tabular}{p{1.8truecm} c c c c c c}
\hline
 & \multicolumn{3}{c}{$<{\rm M}_I>$ = --19.4}
& \multicolumn{3}{c}{$<{\rm M}_I>$ = --20.9} \\
            & $\alpha$ & $\beta$ & $\frac{R_{IBD}}{R_{opt}}$ & $\alpha$ & $\beta$ & $\frac{R_{IBD}}{R_{opt}}$ \\
\hline
VML (s=1.0) &   0.87   &  0.17   & 0.32 &   0.92   &   0.39 & 0.76 \\
VML (s=0.7) &   0.80   &  0.13   & 0.25 &   0.75   &   0.30 & 0.55 \\
VML (s=0.5) &   0.75   &  0.10   & 0.20 &   0.61   &   0.21 & 0.36 \\
\hline
CML     &   0.94   &  0.24   & 0.42 &   1.22   &   0.53 & 1.08 \\
\hline
\hline
 & \multicolumn{3}{c}{$<{\rm M}_I>$ = --22.4} & 
\multicolumn{3}{c}{$<{\rm M}_I>$ = --23.1} \\
            & $\alpha$ & $\beta$ & $\frac{R_{IBD}}{R_{opt}}$ & $\alpha$ & $\beta$ & $\frac{R_{IBD}}{R_{opt}}$ \\
\hline
VML (s=1.0) &   1.28   &  0.64   & 1.34          &   1.29   &   0.69 & 1.53 \\
VML (s=0.7) &   0.94   &  0.52   & 1.05          &   0.93   &   0.57 & 1.18 \\
VML (s=0.5) &   0.75   &  0.42   & 0.81          &   0.75   &   0.46 & 0.91 \\
\hline
CML     &   1.63   &  0.80   & 1.93          &   1.71   &   1.85 & $>$2\\
\hline
\end{tabular}
\end{table}
 
First of all we notice that, when also considering the fitting uncertainties,  
the VML stellar disc contribution to the total RC is different from the CML 
disc contribution. At its peak, the amplitude is lower by $\sim 20\%$.
Noticeably, the VML discs (in combination with the PSS96 DM halo profiles) 
reproduce the kinematic data in an excellent way: the resulting 
$\chi^2$ values are not higher than those of the CML models in  PSS96;
that is to say, the VML hypothesis is by no means excluded by the data. 
With respect to the standard CML decomposition, the overall mass distribution 
of spiral galaxies inferred in the VML scenario is not drastically modified, 
but some major differences are present. These emerge beyond the 
uncertainties on the structural parameters, but rely of course 
of the fundamental assumption that Eq.~6 holds for most spiral galaxies.  
 
The VML disc mass distribution has a higher percentage of its total mass
residing in the inner regions (i.e.\ 70\% within $R< 2\, h_I$,
Fig.~\ref{fig:newMdisc}). The disc normalization, constrained by kinematic 
data, is lower and the inferred total disc mass and {\mbox{M$_\star$/L}} ratio 
are about 1.5 times lower than in the CML case.

In low--luminosity spiral galaxies (Fig.~\ref{fig:decomp}a) dark matter already
dominates the mass distribution at small radii, so that fitting with
VML discs does not influence the inferred halo properties.
At higher luminosity, however, the role of the disc becomes more prominent. 
Since the disc normalization is lower and the VML RC, after its peak 
at $R \sim 1.7 \, h$, falls off more rapidly than the CML RC, in the region 
$R>0.5 \, R_{opt}$ the dark halo velocity contribution must rise more steeply, 
to match the (typically still rising) rotation curve.
 
For galaxies of M$_I$=--21 and brighter, the PSS96 mass models showed a region
of inner baryon dominance (IBD). Defining the corresponding radius $R_{IBD}$
as the region where the baryons contribute more than 50\% of the observed 
circular velocity, the CML mass models yield $R_{IBD}$ as wide as
$R_{opt}$ or more (see Table~1).  Mass models based on VML discs + PSS96 dark 
haloes also show 
an IBD region, but its size is smaller:  $R_{IBD}$ extends beyond $R_{opt}$ 
only in objects of highest luminosity (Fig.~\ref{fig:decomp}b,c,d).
In the luminosity bin of Milky Way--like galaxies 
(M$_I$=--22.3, Flynn et~al.\ 2006), the luminous and DM contribution 
to the rotation curve at $R_{opt}$ are comparable, 
rather than still disc--dominated, as is the case for the CML mass models.
 
We find, as in PSS96, that the  parameters $\alpha$, $\beta$ and galaxy 
luminosity are closely related, so we expect that scaling laws relating
disc and halo masses, luminosity, disc scalelengths, and other galaxy global 
quantities, still hold in the VML scenario. However, we do not attempt
to give explicit fitting formulae (similar to those in PSS96) because
the differences we find with the PSS96 results are moderate. Moreover, 
the VML scenario needs yet to be confirmed by combined 
kinematic/chemo-photometric analysis of {\it individual} objects ---
where scatter is large (Fig.~\ref{fig:colprofiles_T} 
and~\ref{fig:colprofiles_models}).

For galaxies in the highest luminosity bin, there may be 
some weak feature in the co-added RC and in individual RCs that may be traced
back to a  VML disc. At 1--2~$h_I$, the CML +halo rotation curve slightly 
overpredicts $V(R)$ at a $1 \sigma $ level in 2--3 velocity data. The 
VML+halo model instead matches all the data (see Fig.~6 of PSS96 and 
Fig.~\ref{fig:decomp}c,d). Moreover, at $R \sim 3 h$, VML+halo model RCs
tend to remain flat, rather than decline like the CML+halo curves
(Fig.~\ref{fig:decomp}c,d). This is not strongly excluded by the 
co-added curves (due to the larger error bars beyond $R_{opt}$) and could be 
in line with the RC profile of {\it some} fast rotators 
(Spekkens \& Giovanelli 2006). 
 
The investigation of the  core vs.\ cusp issue in spiral galaxies 
requires the  mass modelling of individual RCs; however it is worth 
discussing whether, with VML discs, the NFW haloes may become compatible 
with the observed kinematics.  
Mass modelling of co-added RCs with a CML disc + the ``neutral'' halo of 
Eq.~7 yields a core in the DM density profile; such core is similar to that 
found by the (much more decisive) analysis of the kinematics of selected 
individual objects.
By replacing the CML disc with a VML one, we still find almost the same 
density core; although its size is about $30\% $ smaller, it is still 
large enough to exclude the NFW profile beyond the uncertainties in the 
fitting method. Thus, at face value, VML discs cannot reconcile N-body 
predictions and observations. 

Furthermore, we can resort to the ingenious proof of incompatibility with
the observed kinematics given by Salucci (2001), to show that VML discs 
are more of a problem than a solution for NFW halos.
The argument involves comparing the slope of the 
observed RC's with  the values  predicted  by  the NFW profile.
We here apply this line of reasoning at the radius where 
the disc RC has its maximum, rather than at $R_{opt}$ as in Salucci (2001).
The following chain of facts and evidence 
leads to  the inevitable conclusion that the DM distribution, at least 
in low luminosity spiral galaxies, is shallower than that predicted by 
a NFW halo:
\begin{description}
\item[i)]
At the  radius  in which the disc contribution peaks ($2.2~h$ for CML discs) 
and the {\it disc}  RC has a flat slope, we have
\[ {\frac{d \log V_h}{d \log R}}|_{2.2~h} = 
(1-\beta_{2.2})^{-1} \, \frac{d \log V}{d \log R}|_{2.2~h} \] 
with $\beta_{2.2}=(V_d^2/V^2)|_{2.2 h}$ the fractional disc contribution
to $V^2$ at $2.2 \, h$.
\item[ii)]
A generous estimate is  $\beta_{2.2} \geq 0.15$.
\item[iii)]
For NFW haloes it is  $d \log V_{NFW}/d \log R~ < 0.2$ for $R> 1.5 h$.  
\item[iv)]
For  RCs of low luminosity objects ($M_I<-21$), we have   
$d \log V/d \log R|_{2.2 h} >0.3$ (PSS96).
\end{description}
In the VML mass modelling, the same chain of arguments also 
applies, with only the difference that now the peak radius is at inner 
physical radii, i.e.\ at  $\sim 1.7 h $. Noticeably, at this radius 
the observed RCs show an even steeper slope,
$d \log V/d \log R|_{1.7 h} > d \log V/d \log R|_ {2.2 h}$, and therefore 
the discrepancy  is even larger  than in the  CML case. For instance, 
for objects of $M_I \sim -21$, the observed 
$d \log V/d \log R|_{2.2~h} \sim 0.2$: their RCs at 2.2~$h$ can be, 
in the CML scenario, marginally reproduced by a NFW halo. 
But in the VML case where the disc contribution peaks, we have 
$d \log V/d \log R|_ {1.7~h} \simeq  0.25$, inconsistent with the NFW halo.


\section{Summary and conclusions}
We have investigated the radial profile of the stellar mass-to-light ratio 
in spiral galaxies, and seen how their inclusion changes the picture of the 
mass models of spiral galaxies.   
We derived the  radial variations of {\mbox{M$_\star$/L}} in galactic 
discs, as predicted by chemo--photometric models in the ``inside--out'' 
formation scenario and  as suggested by the observed colour gradients of spiral 
galaxies.
The {\it differential} radial variation in {\mbox{M$_\star$/L}} we predict 
for a ``typical average'' spiral turns out to be largely independent 
of the assumed 
SF history, stellar IMF, global galaxy colour, and other modelling details. 
It follows a profile of the type
\[
\frac{\frac{M_\star}{L_I}(R)}{\frac{M_\star}{L_I}(h_I)} = \exp \left\{ - a(s) 
\left[ \left( \frac{R}{h_I} \right)^s -1 \right] \right\} 
\]
with $0.5 \leq s \leq 1.1$ and $a(s)$ given in Eq.~3b.
The predicted radial {\mbox{M$_\star$/L}} variation  can be  significant,
for instance in the  $I$ band, {\mbox{M$_\star$/L$_I$}} can 
decrease  by up to a factor of two between 1 scalelength and the outskirts of
the stellar disc ($4 h$). We provide  the corresponding variable 
mass--to--light rotation curve $V_{VML}(R)$, that describes our best guess 
of the stellar disc contribution to the circular velocity in 
a "typical average" spiral  galaxy. 
Since small variations ($ \sim 20\%$) in the latter quantity can also affect
the  mass modelling, VML effects should not be neglected in objects 
showing radial colour gradients.

The resulting disc circular velocity $V_{VML}(R)$ has some interesting features
with respect to the classic constant mass--to--light (CML) Freeman curve.
Since {\mbox{M$_\star$/L}} decreases outwards, $V_{VML}(R)$ peaks at smaller 
radii: at 1.5--1.9~$h_I$ (rather than at 2.2~$h$), 
and it approaches the Keplerian fall--off earlier, at 3--3.5~$h_I$
(rather than at $>4~h$). For a given observed RC, the steeper decline 
of the VML stellar contribution at the edge of the optical disc, with respect 
to the CML case, implies there an even steeper radial rise in the halo 
contribution. 
      
With the average VML RC of stellar discs (combined with a cored dark 
matter halo), we have successfully fitted the co-added rotation curves of
spiral galaxies of PSS96. The radial variations in the stellar 
{\mbox{M$_\star$/L}} predicted by chemo--photometric models seem compatible 
with presently available kinematics (Fig.~\ref{fig:decomp}). 
Independent support may come from Salucci et~al.\
(2008), who compared photometric and CML kinematic estimates of
disc masses: despite the generally good agreement (the r.m.s. and the offset 
being only of 0.2~dex and 0.1~dex respectively), there were a few cases 
of photometric masses higher than 
the kinematic ones. These discrepancies could stem from VML discs.
   
The VML disc scenario may be very relevant in two open issues of galaxy 
structure.
Firstly, we have shown that the radial variations of {\mbox{M$_\star$/L}}
allowed by chemo-photometric arguments, cannot reconcile the DM halo density 
profiles derived by mass modelling with those emerging from $\Lambda$CDM 
simulations.
In contrast, the ``standard'' assumption of constant
${\mbox{M$_\star$/L}}$ helps NFW haloes to fit the observed kinematics. 
Because the colours of spiral galaxies typically redden inwards, 
the corresponding ${\mbox{M$_\star$/L}}$ variations are an additional problem 
for NFW profiles.
   
Secondly, we discuss whether, {\it inside the optical disc}, the stars 
alone, without any dark component, could account for the observed 
RCs via a radial variation of {\mbox{M$_\star$/L}}. 
This question was raised in the 80s, when evidence for DM on galactic 
scales started to gather. It is usually bypassed in view of the existence 
of non Keplerian, very extended HI RCs implying a huge mass discrepancy 
beyond the optical region. Nevertheless, the issue of the RC within the optical
region is far from trivial, as no--DM scenarios like MOND and $f(R)$ theories 
of gravity are becoming popular (e.g.\ Sanders \& McGaugh 2002). 

Figure~\ref{fig:noDM} shows the co-added RC of galaxies with 
$<{\rm M}_I>=-19.4$ 
out to $3.5 \, h_I$. It  has a  steep slope  which implies a clear 
mass discrepancy  from $R=h_I$ outwards. We recall that at these luminosities 
the HI disc is dynamically important only outside the region we are considering
here (de Blok et~al.\ 2008). 
In Fig.~\ref{fig:noDM} we show that the co-added curve can be reproduced well 
without  any  DM, with just  a VML  stellar disc. To reach   
this, however, we must invoke an {\it increase} 
in {\mbox{M$_\star$/L}} with radius at the pace of $1+0.35 (R/h_I)^2$.
The corresponding colour variation (PST04) is a {\it reddening } of over 
1~mag in $(B-I)$ between $R=0.5 \, h_I $ and $3 \, h_I$. 
This is totally at odds with the observed radial {\it bluing} of about 
0.4~mag over the same radial range (Fig.~\ref{fig:colprofiles_models}).
We conclude that a dark component must also be present
inside the luminous regions of disc galaxies.

\begin{figure}
\centering
\vskip -2.6truecm
\includegraphics[width=8.5truecm]{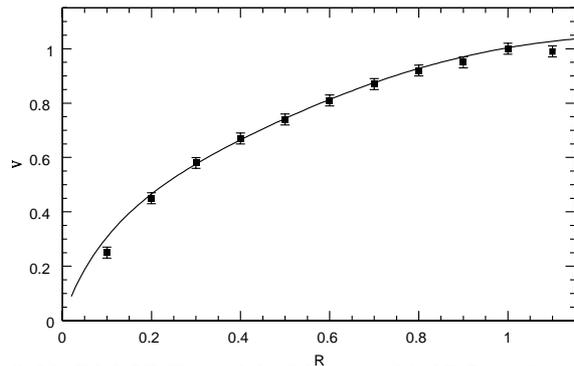}
\vskip -1.truecm
\caption{No DM, VML model of the co--added RC at $M_I=-19.4$. 
The radial coordinate is in units of $R_{opt}=3.2 \, h_I$.}
\label{fig:noDM}
\end{figure}

The existence of  VML discs may play
an important  role in the mass modelling within the MOND paradigm, 
where the observed distribution of the  ordinary matter, stars, and HI gas 
is argued  to  completely account for the observed kinematics. In MOND, 
the circular velocity due to the distribution of the baryonic matter is 
"boosted" with respect to what we would have in Newtonian dynamics. 
{From} about
1 scalelength outward, a MOND self--gravitating Freeman disc 
implies a rotation curve constant with radius and with an amplitude
that, at $R_{opt}$, is about 20\% more than in the Newtonian case.
We consider  high--luminosity  spiral galaxies with
$V_{opt} \sim 250$~km/s, $M_I \sim -23.5$,  $h_I \sim 5.5$~kpc,
belonging to  the second highest velocity bin of the co-added curves in 
PSS96 and shown in Fig.~\ref{fig:MOND}. In these objects 
the dynamical contribution of the HI disc and the stellar bulge are negligible 
in the region between 2 and 3  disc scalelengths. 
For a VML disc the stellar surface density is given by Eq.~5.
We model the RC of a VML disc in MOND dynamics with $s=1$ (shallow profile)
or $s=0.5$ (concave profile). 
Let us recall that in Newtonian gravity both VML discs, or a CML disc,
can fit this same RC when combined with a suitable cored dark halo
(Fig.~\ref{fig:decomp}d).
In MOND, instead, the outcome is  different:
while the $s=1$ VML disc accounts for the  RC, the $s=0.5$ case does not 
(Fig.~\ref{fig:MOND}).
This result is not a test of MOND, since we do not know the average $s$ in 
spiral galaxies, but indicates that in alternative theories to dark matter,
the effects of {\mbox{M$_\star$/L}} variations are much greater and potentially
testable.

In the disc+dark halo mass modelling, the CML and VML discs yield equally good 
fits to the observed RCs (Fig.~\ref{fig:decomp}), so that one cannot use 
the RC to significantly reveal VML features. This is a non--trivial result: 
it is due not
to an intrinsic degeneracy of the mass modelling, but to the fact that
the allowed {\mbox{M$_\star$/L}} variations are not very large.
In the ``average'' late type spiral galaxy, chemo--photometric arguments
indicate a decrease in {\mbox{M$_\star$/L$_I$}} by a factor of 2
over the optical region, which is hardly distinguished from the kinematics.
In the simplest VML disc case ($s=1$ in Eq.~\ref{eq:Sigma_profile}), for 
instance, the surface density distribution remains exponential, with just a 
scalelength about 20\% shorter than that of the $I$--band luminosity profile.

Noticeably, in alternative scenarios to DM where the kinematics are 
completely determined by the distribution of the baryonic matter,
the mild radial {\mbox{M$_\star$/L}} gradients we predict are expected 
to play a relevant role.

Although the predicted {\mbox{M$_\star$/L}} gradients do not radically alter
the main results of traditional mass modelling, our comparison of the VML 
and CML assumptions shows that VML discs have an impact on RC decomposition: 
the trend is to obtain smaller disc normalization 
and smaller, but non--negligible, dark halo cores. The NFW halo profile
becomes even less compatible with the observed kinematics in the inner regions;
therefore, \mbox{M$_\star$/L}} gradients should in general not be neglected 
in mass modelling, and a combined chemo--photometric and dynamical analysis 
of individual objects (especially those showing strong colour gradients) 
is the best approach.

More ambitious, but not beyond reach, is the idea of using detailed 
kinematic analysis as an independent test of the predictions 
of chemo--photometric models. By turning the above argument around and 
by assuming a complete inner baryon dominance in the innermost regions, 
one can possibly aim at measuring the variations in the stellar 
{\mbox{M$_\star$/L}} directly from high--quality rotation curves and 
check the predictions of chemo-photometric models. Also, since the prediction
is that the scalelength of the stellar density profile is shorter than
the luminous scalelength (by at least 20\%, $s=1$ case), future 
investigations could aim at revealing VML discs by estimating the 
dynamical scalelengths from the observed RCs and then comparing them with 
the photometric ones.
 
\begin{figure}
\centering
\vskip -2.6truecm
\includegraphics[width=8.5truecm]{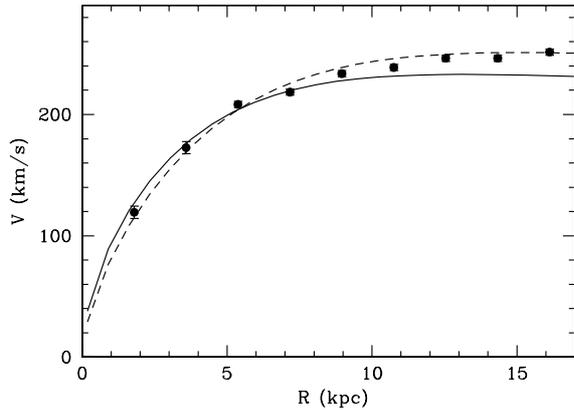}
\vskip -1.truecm
\caption{MOND VML mass model of a co-added rotation curves of
luminous objects.  $s=1$ (dashed) and $s=0.5$ (solid) cases.}
\label{fig:MOND}
\end{figure}


\begin{acknowledgements}
We thank the anonymous referee for many detailed comments that helped us
improve the presentation.
We are grateful to Chris Flynn for careful reading and useful remarks on the
manuscript. LP acknowledges kind hospitality from SISSA and from the Department
of Astronomy in Padua on various visits.
This study was financed by the Academy of Finland (grant nr.~208792), by
the 6$^{th}$ Framework Programme of the European Commission (Marie Curie Action
nr.~MEIF-CT-2005-010884) and by the Italian MIUR.  We used the VizieR On--line
Data Catalogue to retrieve observational data used in this paper.
\end{acknowledgements}
 


\begin{thebibliography}{}
\bibitem{}
Ashman K.M., Salucci P., Persic M., 1993, MNRAS 260, 610
\bibitem{}
Bakos J., Trujillo I., Pohlen M., 2008, ApJ 683, L103
\bibitem{}
Bell E.F., Bower R.G., 2000, MNRAS 319, 235
\bibitem{}
Bell E.F., de Jong R.S., 2000, MNRAS 312, 497
\bibitem{}
Bell E.F., de Jong R.S., 2001, ApJ 550, 212 
\bibitem{}
de Blok W.J.G., 2007, in Island Universes. Structure and evolution of disc
galaxies, ed.\ R.S.\ de Jong, ASSP, Springer, p.~89
\bibitem{}
de Blok W.J.G., Bosma A., 2002, A\&A 385, 816
\bibitem{}
de Blok W.J.G., Walter F., Brinks E., Trachternach C., Oh S.H., Kennicutt R.C.,
2008, AJ 136, 2648
\bibitem{}
Boissier S., Prantzos N., 1999, MNRAS 307, 857
\bibitem{}
Boissier S., Prantzos N., 2000, MNRAS 312, 398 
\bibitem{}
Boissier S., Prantzos N., 2001, MNRAS 325, 321
\bibitem{}
Burkert A., 1995, ApJ 447, L25
\bibitem{}
Burkert A., Truran J.W., Hensler G., 1992, ApJ 391, 651
\bibitem{}
Chabrier G., 2001, ApJ 554, 1274
\bibitem{}
Chabrier G., 2002, ApJ 567, 304
\bibitem{}
Chiappini C., Matteucci F.M., Gratton R., 1997, ApJ 477, 765
\bibitem{}
Chiappini C., Matteucci F., Romano D., 2001, ApJ 554, 1044
\bibitem{}
Dopita M.A., Ryder S.D., 1994, ApJ 430, 163
\bibitem{}
Flynn C., Holmberg J., Portinari L., Fuchs B., Jahrei\ss\ H., 2006, MNRAS
372, 1149
\bibitem{}
Freeman K., 1970, ApJ 160, 811
\bibitem{}
Garnett D., Shields G., Skillman E., Sagan S., Dufour R., 1997, ApJ 489, 63
\bibitem{}
Gentile G., Salucci P., Klein U., Vergani D., Kalberla P., 2004, MNRAS 351 903
\bibitem{}
Gentile G., Burkert A., Salucci P., Klein U., Walter F., 2005, ApJ 634, L145
\bibitem{}
Gummersbach C.A., Kaufer A., Sch\"afer D.R., Szeifert T., Wolf B., 1998, 
A\&A 338, 881
\bibitem{}
Jansen R.A., Franx M., Fabricant D., Caldwell N., 2000, ApJS 126, 271
\bibitem{}
de Jong R.S., 1996a, A\&AS 118, 557
\bibitem{}
de Jong R.S., 1996b, A\&A 313, 45
\bibitem{}
de Jong R.S., 1996c, A\&A 313, 377
\bibitem{}
de Jong R.S., van der Kruit P.C., 1994, A\&AS 106, 451
\bibitem{}
Kassin S., de Jong R.S., Weiner B.J., 2006, ApJ 643, 804
\bibitem{}
Kennicutt R.C., 1983, ApJ 272, 54
\bibitem{}
Kennicutt R.C., Tamblyn P., Congdon C.E., 1994, ApJ 435, 22
\bibitem{}
Kent S.M., 1986, AJ 91, 1301
\bibitem{}
Kranz T., Slyz A., Rix H.-W., 2003, ApJ 586, 143 
\bibitem{}
Kroupa P., 1998, ASP Conf.\ Ser.\ 134, 483
\bibitem{}
Kuzio de Naray R., McGaugh S.S., de Blok W.J.G., Bosma A., 2006, 
ApJS 165, 461
\bibitem{}
Larson R.B., 1976, MNRAS 176, 31
\bibitem{}
Larson R.B., 1998, MNRAS 301, 569
\bibitem{}
L\'opez--Corredoira M., Cabrera--Lavers A., Mahoney T.J., Hammersley P.L., 
Garz\'on F., Gonz\'ales--Fern\'andez C., 2007, AJ 133, 2007
\bibitem{}
Lynden--Bell D., 1975, Vistas in Astr.\ 19, 229
\bibitem{}
Martin P., Roy J.R., 1994, ApJ 424, 599
\bibitem{}
Matteucci F., Fran\c cois P., 1989, MNRAS 239, 885
\bibitem{}
Moll\'a M., Diaz A.I., 2005, MNRAS 358, 521
\bibitem{}
Moll\'a M., Ferrini F., Diaz A.I., 1996, ApJ 466, 668
\bibitem{}
Moll\'a M., Ferrini F., Diaz A.I., 1997, ApJ 475, 519
\bibitem{}
Navarro J.F., Frenk C.S., White S.D.M., 1996, ApJ 462, 563
\bibitem{}
Navarro J.F., Frenk C.S., White S.D.M., 1997, ApJ 490, 493
\bibitem{}
Pagel B.E.J., 1997, Nucleosynthesis and Chemical Evolution of Galaxies,
Cambridge University Press
\bibitem{}
Persic M., Salucci P., 1988, MNRAS 234, 131
 \bibitem{}
Persic M., Salucci P., Stel F., 1996, MNRAS 281, 27 (PSS96) 
\bibitem{}
Portinari L., Chiosi C., Bressan A., 1998, A\&A 334, 505
\bibitem{}
Portinari L., Chiosi C., 1999, A\&A 350, 827
\bibitem{}
Portinari L., Chiosi C., 2000, A\&A 355, 929
\bibitem{}
Portinari L., Sommer--Larsen J., Tantalo R., 2004, MNRAS 347, 691 (PST04)
\bibitem{}
Prantzos N., Boissier S., 2000, MNRAS 313, 338
\bibitem{}
Press W.H., Teukolsky S.A., Vetterling W.T., Flannery B.P., 1992, Numerical
Recipes, Cambridge University Press
\bibitem{}
Renda A., Kawata D., Fenner Y., Gibson B.K., 2005, MNRAS 356, 1071
\bibitem{}
Roberts M.S., Haynes M.P., 1994, ARA\&A 32, 115
\bibitem{}
Ro\v{s}kar R., De Battista V.P., Quinn T.R., Stinson G.S., Wadsley J., 2008,
ApJ 684, L79
\bibitem{}
Ryder S.D., Dopita M.A., 1994, ApJ 430, 142
\bibitem{}
Salpeter E.E., 1955, ApJ 121, 161
\bibitem{}
Salucci P., 2001, MNRAS 320, L1
\bibitem{}
Salucci P., Yegorova I., Drory N., 2008, MNRAS 388, 159
\bibitem{}
Salucci P., Walter F., Borriello A., 2003, A\&A 409, 53
\bibitem{}
Salucci P., Lapi A., Tonini C., Gentile G., Yegorova I., Klein U., 2007,
MNRAS 378, 41
\bibitem{}
Sanders R.H., McGaugh S.S., 2002, ARA\&A 40, 263
\bibitem{}
Sch\"onrich R., Binney J., 2009, MNRAS in press (arXiv:0809.30006)
\bibitem{}
Shaver P.A., McGee R.X., Newton L.M., Danks A.C., Pottasch S.R., 1983, 
MNRAS 204, 53
\bibitem{}
Smartt S.J., Rolleston W.R.J., 1997, ApJ 481, L47 
\bibitem{}
Sommer--Larsen J., 1991, MNRAS 250, 356
\bibitem{}
Sommer--Larsen J., G\"otz M., Portinari L., 2003, ApJ 596, 47
\bibitem{}
Spano M., Marcelin M., Amram P., Carignan C., Epinat B., Hernandez O., 2008,
MNRAS 383, 297
\bibitem{}
Spekkens K. Giovanelli R., 2006, AJ 132, 1426
\bibitem{}
Tinsley B., 1980, Fund.\ Cosmic Phys.\ 5, 287
\bibitem{}
Treuthardt P., Salo H., Buta R., 2008, AJ in press (arXiv:0809.2478)
\bibitem{}
Verheijen M.A.W., 1997, PhD thesis, University of Groningen, The Netherlands
\bibitem{}
van Zee L., Salzer J.J., Haynes M.P., O'Donoghue A.A., Balonek T.J., 1998,
AJ 116, 2805
\end{thebibliography}
\end{document}